\documentclass[12pt]{article}
\usepackage{epsfig}

\begin{document}

\title{Tearing transition and plastic flow in superconducting thin films}

\author{\bf M.-Carmen Miguel$^1$ \& Stefano Zapperi$^2$}
\maketitle
\vskip -1cm 
\centerline{$^1$Departament de F\'{\i}sica
Fonamental, Facultat de F\'{\i}sica, Universitat de Barcelona,} 
\centerline{Diagonal 647, E-08028, Barcelona, Spain} 
\centerline{$^2$INFM unit\`a di Roma 1 and SMC, Dipartimento di Fisica,
Universit\`a "La Sapienza",} \centerline{P.le A. Moro 2, 00185 Roma, Italy}

\vspace{1truecm}

\noindent A new class of artificial atoms, such as synthetic
nanocrystals or vortices in superconductors, naturally self-assemble
into ordered arrays. This property makes them applicable to the
design of novel solids, and devices whose properties often depend on
the response of such assemblies to the action of external
forces. Here we study the transport properties of a vortex array in
the Corbino disk geometry by numerical simulations. In response to an
injected current in the superconductor, the global resistance
associated to vortex motion exhibits sharp jumps at two threshold
current values. The first corresponds to a tearing transition from
rigid rotation to plastic flow, due to the reiterative nucleation
around the disk center of neutral dislocation pairs that unbind and
glide across the entire disk. After the second jump, we observe a
smoother plastic phase proceeding from the coherent glide of a larger
number of dislocations arranged into radial grain boundaries.

\vspace{1truecm}

\noindent The production of ordered self-assembled structures of
various materials as diverse as synthetic nanocrystals, magnetic
colloids, charged particles in Coulomb crystals, proteins and
surfactants, or vortices in type II superconductors and in
Bose-Einstein condensates, has attracted much interest for various
fundamental and practical reasons which are ultimately concerned with
their collective properties (optical, magnetic, mechanical, or
transport properties)~\cite{MUR-00,MIT-01,PER-01}. In particular, much
experimental and theoretical effort has been devoted to characterizing
the phase diagram of type II superconductors~\cite{BRA-95}. Depending
on the value of the magnetic field $H$, temperature $T$, and sample
preparation, vortices can either form a crystal~\cite{ABR-57}, which
at higher temperatures melts into a liquid
~\cite{SAF-92,AVR-01,BOC-02,SCH-02}, or, due to quenched disorder,
they can be found in more complex phases, such as the vortex
glass~\cite{FIS-91}, the Bose glass~\cite{NEL-00} or the Bragg glass
~\cite{GIA-95,KLE-01}.  Of special importance is the non-equilibrium
response of vortex matter to the flow of an external current
~\cite{LED-98,CRA-99}, because the dissipative motion of the vortices
induces an undesirable macroscopic resistance.  The moving phase can
be as simple as the collective motion of an elastically deforming
vortex crystal, or can be more complex, such as in plastic
~\cite{BHA-93} or in channel vortex flow~\cite{MAR-97}.

Transport experiments in superconductors are often performed in a
strip geometry: current is injected in one side and removed from the
opposite side~\cite{BHA-93,MAR-97}. The current-voltage (I-V) curve
provides an indirect measure of vortex dynamics, because vortex motion
induces an electric field proportional to the vortex velocities.  It
has been noticed that the sample boundary has an important effect on
the moving vortex phase complicating the interpretation of the results
~\cite{PAL-00,PAL-00b}; a problem that is overridden in the Corbino disk
geometry~\cite{CRA-99,PAL-00b,DAN-96,LOP-99}.  There the current is
applied at the disk center and flows radially towards the boundary.
Vortices tend to move in concentric circles without crossing the
sample boundaries, avoiding edge contamination.

Vortices in the Corbino geometry experience a force
gradient~\cite{CRA-99}, and thus exhibit intriguing dynamic phases as
a function of $T$, $H$, and $I$. The vortex velocity profiles have
been evaluated after measuring the voltage drop across a series of
contacts placed radially on a YBa$_2$Cu$_3$O$_{7-\delta}$
disk~\cite{LOP-99}. For low currents and temperatures, all the
vortices were found to move as a rigid solid, giving rise to a linear
velocity profile $v(r)=\Omega r$, where $r$ is the distance from the
disk center. Above a threshold current $I_0$, the vortex crystal
cannot sustain the shear stress induced by the resulting inhomogeneous
Lorentz force, and the response becomes plastic.  Finally, above the
vortex lattice melting temperature $T_M$, the velocity profile is
fluid-like and decays as $v(r)\sim 1/r$. A theoretical model of vortex
flow in the Corbino geometry has been analyzed, and shows the shear
yielding as a dislocation unbinding
transition~\cite{MAR-00,BEN-02}. Plastic flow would appear as soon as
the current-induced shear stress is large enough to separate an
existing pair of bound dislocations.

In this paper, we first study transport in the Corbino disk by $T=0$
molecular dynamics (MD) simulations of interacting vortices
\cite{JEN-88,FAN-01,CHE-03}.  As in the experiments, for low currents
we find a linear velocity profile that corresponds to the rigid
rotation of the vortex lattice. Above a threshold current $I_0$, the
profile ceases to be linear, indicating the onset of plastic flow. Our
simulations enable a close inspection of the lattice topology, which
unveils the microscopic origin of this dynamic transition: at and
above $I_0$, {\em new dislocation} pairs are created, mainly within
the highly strained central region, which readily unbind and glide
along all possible crystalline directions giving rise to plastic
flow. These processes occur repeatedly, yielding a strongly
fluctuating voltage noise, which is reminiscent of the intermittent
behavior observed in plastically deforming crystals
\cite{MIT-01,MIG-01}.  For currents larger than a second threshold
$I_1$, we observe that voltage fluctuations decrease and vortices end
up moving in uncorrelated annular channels, displaying a laminar $1/r$
velocity profile. In this regime, we find a larger amount of
dislocations in the crystal, most of them forming radial grain
boundaries that span the entire disk and glide in the tangential
direction. The crystal reorientations associated to the presence of
these grain boundaries make possible a steady regime of plastic
deformation in the azimuthal direction. In addition, the exponential
screening of shear stress produced by grain boundaries~\cite{Hirth92}
enhances the number of nucleation events. The number of dislocations
after the second jump at $I_1$ reaches a maximum value, corresponding
to the presence of quite densely packed grain boundaries. We observe
that the plastic deformation of the crystalline film proceeds with the
glide motion of these grain boundaries.

We measure the plastic threshold current $I_0$ for different values of
the disk radius $D$ and vortex number $N$, and later show (in
Fig.~\ref{fig5}) that the data collapse into a single curve when plotted
against the crystal average lattice spacing $a$. As predicted
previously~\cite{BEN-02},$I_0$ is proportional to the vortex crystal
shear modulus $c_{66}$. Indeed, the curve follows very closely the
dependence of $c_{66}$ on the lattice spacing or field strength
\cite{MIG-00}. A simple evaluation of the energy cost of a new
dislocation pair provides a good quantitative estimate of this
threshold current.

In the Corbino geometry, a disk-shaped superconductor is placed in a
magnetic field parallel to the disk axis, and a current $I$ is
injected at a metal contact in the disk center and removed at the disk
boundary. Thus, the current density inside the disk is given by ${\bf
J}(r)=\hat{r}I/(2\pi r h)$, where $h$ is the thickness of the
specimen.  This radial current generates an azimuthal Lorentz force
acting on the vortices ${\bf f}_L(r)= \hat{\theta} \Phi_0 J(r) /c$,
where $\Phi_0$ is the quantized flux carried by the vortices, $c$ is
the speed of light, and $\hat{\theta}$ is the azimuthal versor. In
addition, a pair of vortices interact with each other via a long-range
force ${\bf f}_{vv}({\bf r})=AK_1(|{\bf r}|/\lambda)\hat{r}$, where
$A=\Phi_0^2/(8\pi^2\lambda^3)$, $\lambda$ is the London penetration
length, and $K_1$ is a Bessel function~\cite{degennes}. Taking into
account these interactions, we solve numerically the dynamics of $N$
vortices confined in a disk of radius $D$ (see the Methods section,
describing the technical details of the simulations).

\begin{figure}[ht]
\centerline{\psfig{file=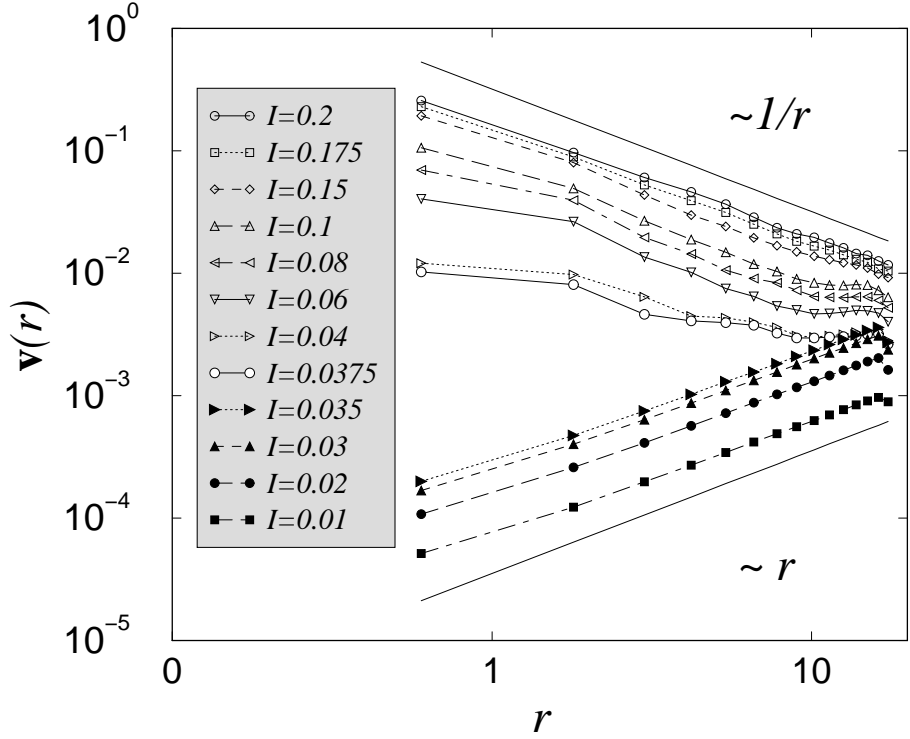,width=12cm,clip=!}}
\caption{The velocity profiles as a function of the applied current
for $N=1032$ vortices in a disk of radius $D=18$.  For $I<I_0=0.036$
the profile is linear, corresponding to a rigid rotation of the
lattice. For $I>I_0$ the profile deforms, indicating plastic flow. At
high drives, the profile simply decays as $1/r$, where $r$ is the
distance from the disk center.}
\label{fig1}
\end{figure}

As in previous experiments~\cite{LOP-99}, we first measure azimuthal
velocities $v$ and compute the variation of the velocity profile
$v(r)$ with $I$ (see Fig.~\ref{fig1}).  At low currents, the velocity
profile follows a linear law, $v(r)=\omega r$ with angular velocity
$\Omega \propto I$. This corresponds to a rigid rotation of the vortex
crystal, as observed experimentally~\cite{LOP-99}. Above a threshold
$I_0$, the velocity profile starts to deform, with large tangential
velocities in the center, which then decay towards the boundary. At
higher currents, the profile becomes smoother, decaying as $1/r$,
characteristic of a laminar response.

\begin{figure}[t]
\centerline{\psfig{file=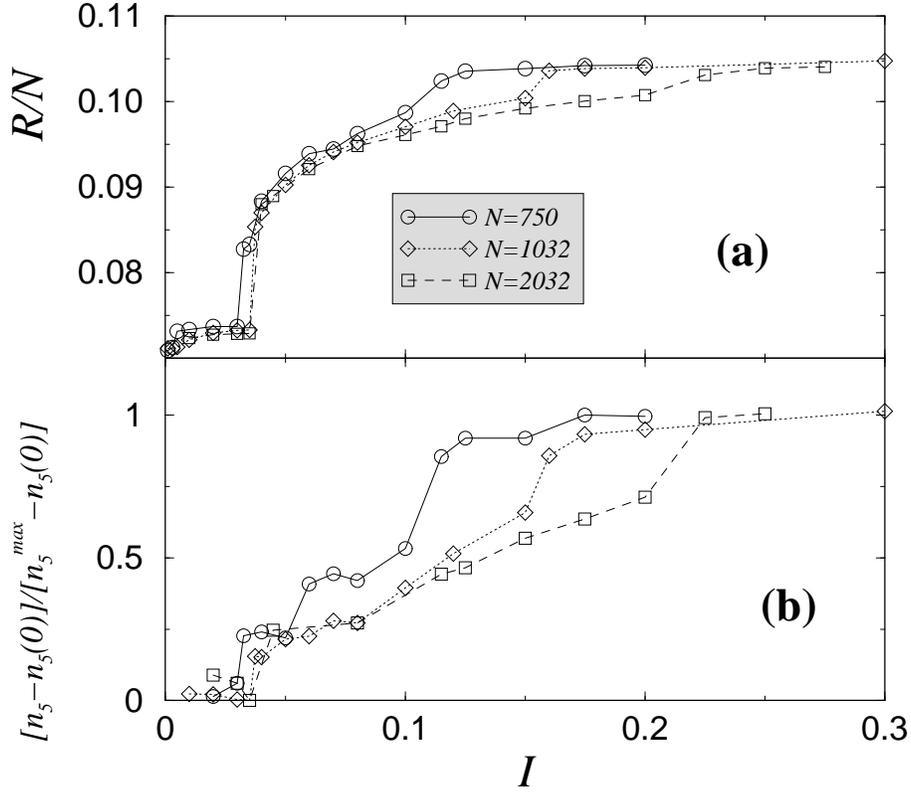,width=12cm,clip=!}}
\caption{The resistance and the number of fivefold vortices as a
function of the current. (a) The normalized resistance $R/N$ as a
function of the applied current for $N$ vortices in a disk of radius
$D=18$. (b) The excursions in the average steady number of fivefold
coordinated vortices $n_5$ for different $N$ by plotting
$(n_5-n_5(0))/(n_5^{max}-n_5(0))$.}
\label{fig2}
\end{figure}

To better identify the transitions in the system rheology, we measure
the variations of the flow resistance $R\equiv \sum_i v_i /I$ with
$I$. After an initial transient, the resistance reaches a steady
state, which fluctuates strongly in the plastic regime and is much
smoother in the solid and laminar phases. Figure \ref{fig2}(a) shows
the steady-state resistance for different values of $N$ (at $D=18$) as
a function of $I$. We have normalized these curves by the
corresponding number of vortices $N$ to better visualize their
characteristic features. The curves show a first sharp jump around
$I_0$ corresponding to the breakdown of the linear velocity profile,
and a smaller jump at $I_1$, indicating the onset of the hyperbolic
profile. The final plateau scales with the number of moving vortices
$N$. Indeed in this laminar regime, a scaling factor of $N/D$ follows
from a simple continuum approximation with a constant density of
vortices. A direct inspection of the topology of the lattice allows
the nature of the transitions to be clarified. We construct the
Delaunay triangulation of the vortex positions in the disk to
characterize their topology.  Most of the vortices are sixfold, as in
a perfect triangular lattice such as the Abrikosov lattice. A pair of
fivefold and sevenfold neighboring vortices identifies an edge
dislocation in the lattice, a topological defect characterized by its
Burgers vector ${\bf b}$ \cite{Hirth92}. Dislocations produce
long-range stress and strain fields in the host crystal, experience
the so-called Peach-Koehler force due to the local stress, and move
mainly by gliding along the direction of ${\bf b}$ \cite{Hirth92}.

\begin{figure}[t]
\centerline{\psfig{file=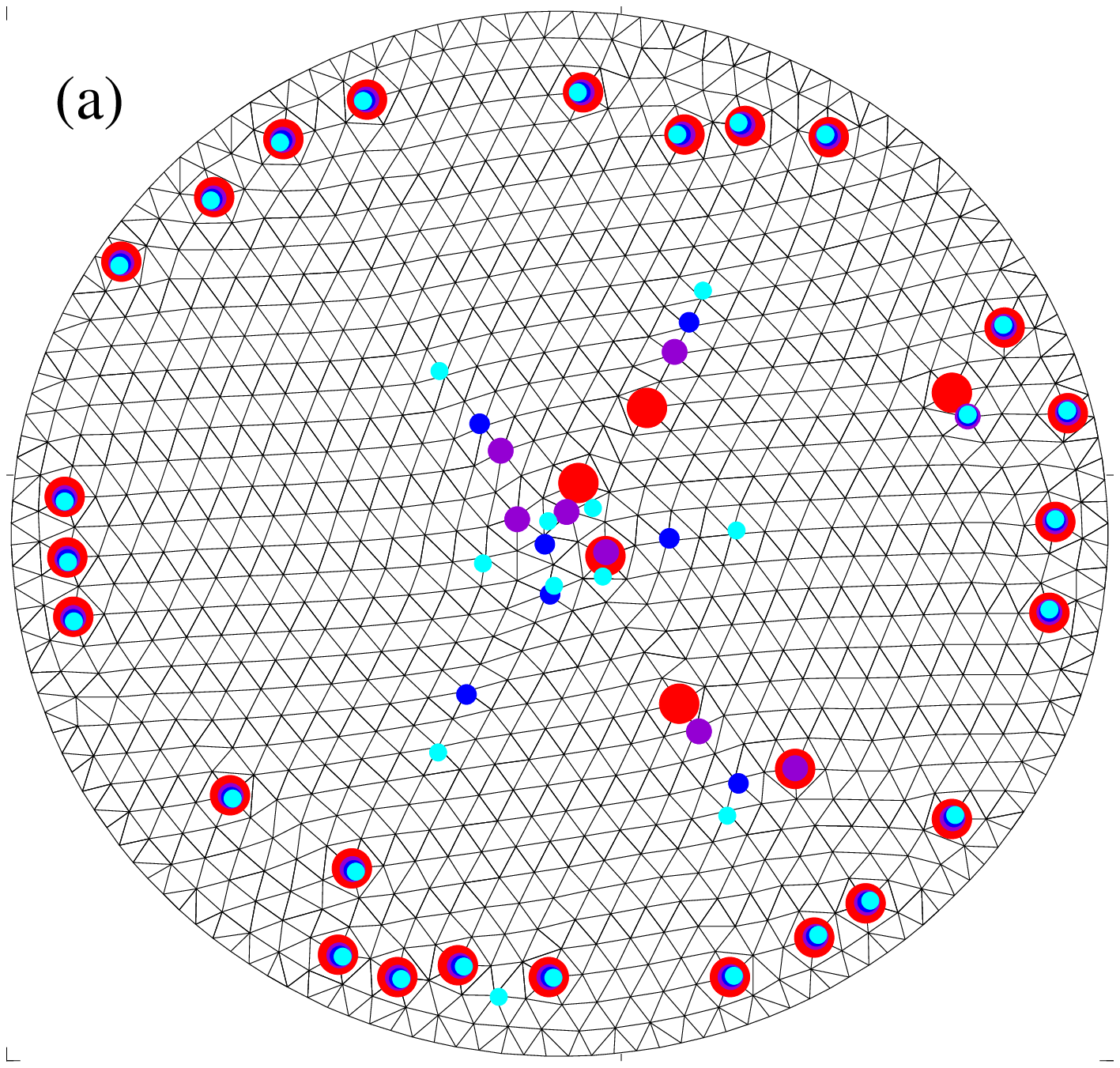,width=10cm,clip=!}}
\centerline{\psfig{file=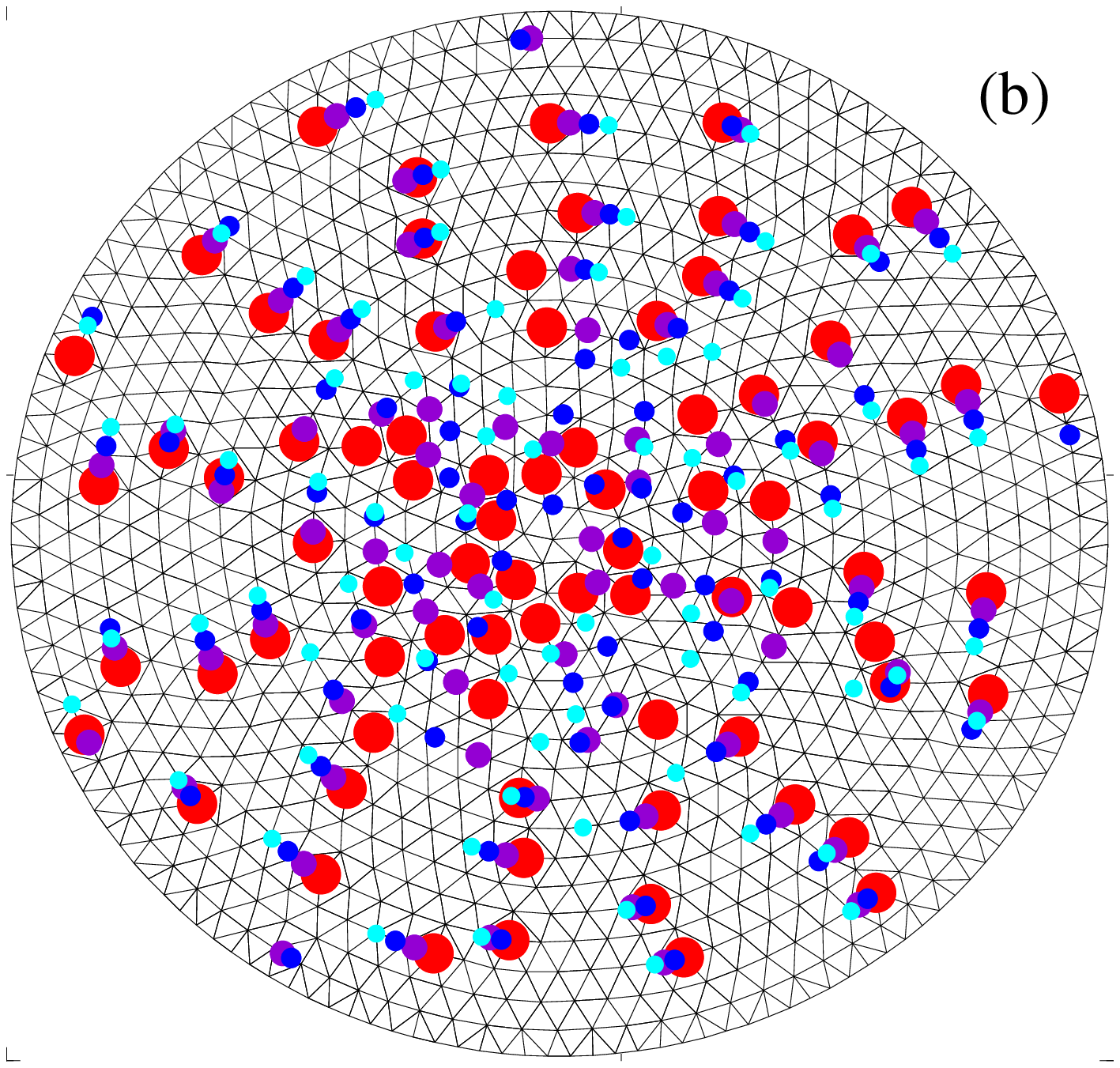,width=10cm,clip=!}}
\caption{A series of snapshots illustrating dislocation dynamics.  The
system consists of $N=1032$ vortices in a disk of radius $D=18$ under
two different applied currents: (a) $I=0.04$, and (b)
$I=0.3$. Fivefold coordinated vortices are highlighted with colored
circles. Different colors correspond to four different time steps
separated by an interval $\Delta t=100$.  Initially they are red, then
violet, blue, and cyan.  As a reference, on each plot we also show the Delaunay
triangulation for the initial time step considered. Notice the radial
motion of some defects in the plastic phase (a), and the tangential
glide of grain boundaries in the laminar phase (b).}
\label{fig3}
\end{figure}

For $I<I_0$, all vortices within the bulk of the disk are six-fold
whereas a large number of fivefold- and sevenfold-coordinated vortices
are only observed along the boundary. These are geometrically
necessary dislocations and disclinations, which need to be present in
order to adjust a triangular lattice into a circular geometry. Like
the resistance, the number of five/sevenfold coordinated vortices
fluctuates around a steady average value after an initial
transient. Figure \ref{fig2}(b) shows the behavior of the average
steady number of fivefold vortices $n_5$ as a function of the
current. We have subtracted the average number of geometrically
necessary boundary defects $n_5(0)$, and normalized the curves by its
maximum value $n_5^{max}-n_5(0)$ to better visualize their main
features. The curves in Fig.~\ref{fig2}(b) closely resemble the
behavior of the resistance in Fig.~\ref{fig2}(a), with jumps at $I_0$
and $I_1$ (see also Supplementary Information, Fig. S1 corresponding
to a disk of radius $D=36$). As the current overcomes $I_0$, new
defects start to nucleate near the center of the lattice. Typically,
we observe the reiterative formation of new dislocation dipoles (two
dislocations with opposite Burgers vectors), that unbind and glide
along the direction of their Burgers vector, in most cases towards the
disk boundary (see Fig.~\ref{fig3}(a) and supplementary movies 1 and
2). To accommodate the shear stress generated by the external current,
the crystal should nucleate dislocations that are able to glide either
radially or tangentially. Nevertheless, in the undistorted triangular
lattice (or when the concentration of free dislocations is low), the
dislocations that are nucleated are the most elementary, with Burgers
vectors along the three basic crystalline directions. Of those, only
the ones with ${\bf b}$ almost parallel to the radial direction can
easily glide over long distances due to the Peach-Koehler forces
involved.

\begin{figure}[t]
\centerline{\psfig{file=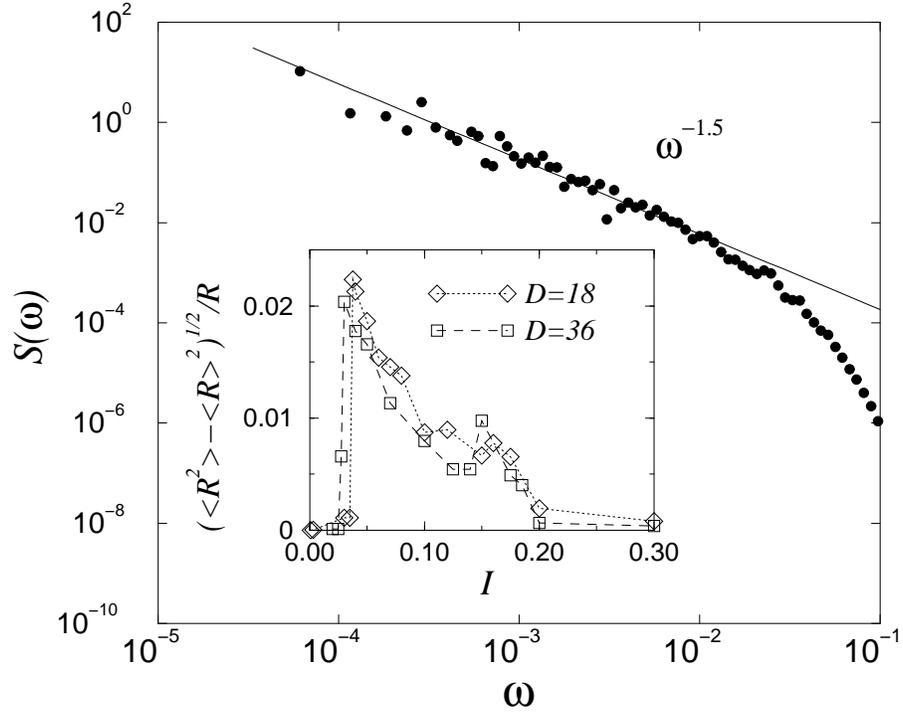,width=12cm,clip=!}}
\caption{The resistance noise power spectrum for $N=1032$ vortices in
a disk of radius $D=18$ under an applied current $I=0.0375$. There is
a region of power law decay well fit by $\omega^{-1.5}$. The inset
shows the relative resistance noise standard deviation as a function
of the current for $D=18$ and $D=36$. The peaks correspond to the
transitions form solid to plastic and from plastic to laminar.}
\label{fig4}
\end{figure}

\begin{figure}[t]
\centerline{\psfig{file=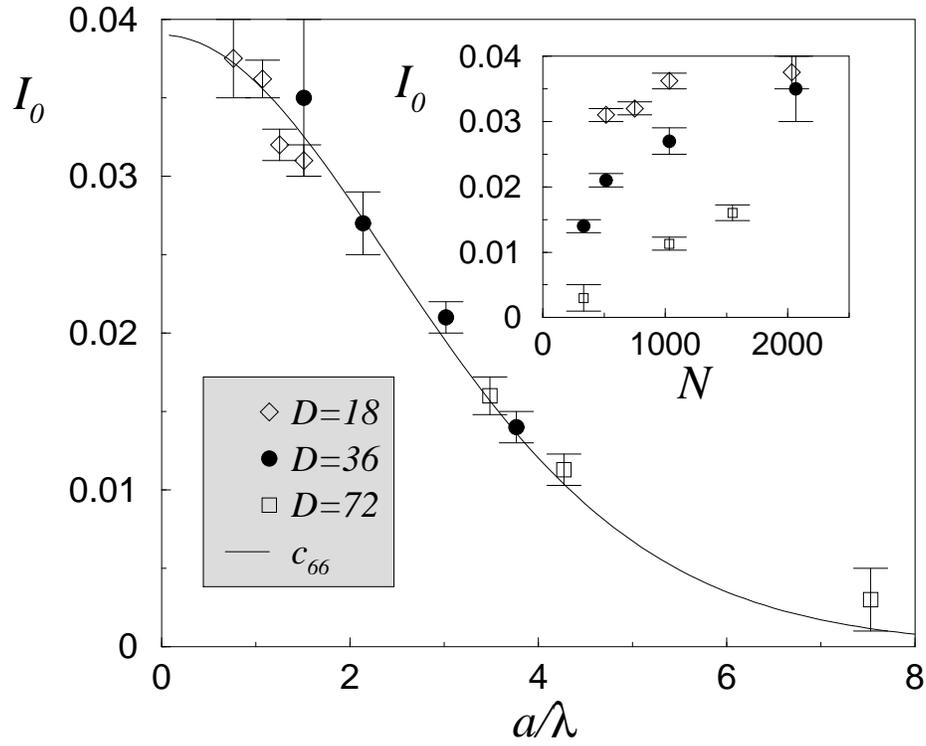,width=12cm,clip=!}}
\caption{The threshold current $I_0$. In the inset we report the value
of $I_0$ as a function of the number of vortices $N$ for different
disk radii $D$. When plotted as a function of the lattice spacing $a$,
the data collapse into a single curve, which follows the decay of
$c_{66}$ (main plot).}
\label{fig5}
\end{figure}

As time goes on, the dislocation flow process exhibits an erratic
character, because dislocation pairs at short distances may annihilate
each other or react to form a new dislocation; they assist the
nucleation process at intermediate distances, and may even form
various metastable structures. This intricate process is reflected by
the strongly fluctuating flow resistance and the non-linear velocity
profile.  Figure~\ref{fig4} shows that at the onset of the plastic
phase, the resistance noise power spectrum $S(\omega)$ (where $\omega$
is the frequency) displays a non-trivial power law decay
$\omega^{-\beta}$ with $\beta=1.5$. This is also apparent from the
inset of Fig.~\ref{fig4}, which displays the relative standard
deviation of the resistance fluctuations as a function of current. We
can then conclude that in the Corbino disk, strong voltage noise is a
fingerprint of the onset of plastic deformation in the lattice.

Further increase of the current beyond $I=I_1$ results in a smoother
flow, with most vortices moving in uncorrelated concentric
trajectories. Here dislocation flow appears to be quite peculiar
because a significant number are settled in walls (grain boundaries)
oriented along the radial direction, which tend to glide coherently in
the azimuthal direction. Grain boundaries produce the necessary
deformations of the crystal that allow a stationary tangential flow
(see Fig.~\ref{fig3}(b) and supplementary movies 3 and 4). Besides,
the long-range stress field generated by free dislocations is screened
out with their arrangement into grain boundaries. The screening length
grows with the average distance separating contiguous dislocations in
the wall, that is, the grain boundary spacing. In a close-packed grain
boundary, the spacing is of the order of a few crystal lattice
constants $a$, and consequently, the number of dislocations in a
radial grain boundary of size $D$ grows as $D/a$. We indeed observe
that the asymptotic number of dislocations in the laminar regime grows
as $n_5^{max}\sim D/a \sim \sqrt{N}$ (see Supplementary Information,
Fig. S2). We also observe that relative fluctuations of dislocation
number and resistance (see inset in Fig.~\ref{fig4}) are consequently
reduced in this regime. Thus at and above $I_1$, the rate of
nucleation of new dislocations is big enough to ensure the formation
and maintenance of these type of grain boundaries whose cooperative
motion marks the new regime of plastic deformation.

Motivated by the observation that the onset of plasticity corresponds
to the reiterative nucleation of topological defects around the center
of the disk, we estimate $I_0$ by computing the energy cost of a
dislocation dipole \cite{BEN-02}. Very much like in any other ordinary
crystal, the energy cost of an edge dislocation in the vortex lattice
is made up of two contributions: the core energy $F_c$ and the elastic
energy $F_e$. The elastic energy cost $F_e$ is proportional to the
shear modulus of the crystal, and grows with the logarithm of the
system size $D$ \cite{Hirth92}. The order of magnitude of the core
energy per unit length can be estimated using the same variational
argument that has been proposed~\cite{Chaikin95} for a vortex line in
the xy-model. It turns out that $F_c$ is proportional to
$c_{66}b^2/(4\pi)$ for each dislocation, where $c_{66}$ is the local
shear modulus of the vortex lattice.

The elastic energy cost of a new dislocation dipole, such as the ones
observed in the plastic phase, is independent of the system size and
grows with the relative distance between the dislocations
$r_d$---which, right at the moment of the creation event, is of the
order of the lattice spacing---as $b^2 c_{66}/(2\pi) \ln(r_d/r_c)$,
where $r_c$ is a variational short distance cutoff. Because both $r_d$
and $r_c$ are of the order of the lattice spacing, the core energy of
the new dislocation dipole $2F_c$ is the leading energetic
contribution for the proliferation of new pairs. The inhomogeneous
elastic shear stress and strain induced by the external current in the
Corbino disk geometry have been calculated analytically in
Ref.\cite{BEN-02}. According to those results, the elastic energy
stored in a region around the disk center of size $R_d$---of the order
of the spatial extent of a dislocation dipole---is roughly equal to $
(R_d IB/4\pi c h)^2 \pi/(4 c_{66})$, where $B=\phi_0 n$ is the average
magnetic induction and $n$ the areal density of flux lines. This
energy is released by the formation of new dislocation pairs. On
balancing the core energy with the elastic energy provided by the
external current one can estimate the transition current
\begin{equation}
I_0 = \frac{4 \sqrt{2} c h b c_{66}}{R_d B}=\frac{4\sqrt{2} c h b
\phi_0}{R_d(8\pi\lambda)^2}\frac{c_{66}}{\tilde{c}_{66}},
\end{equation}
where $\tilde{c}_{66}=B\phi_0/(8\pi\lambda)^2$ is the long wavelength
shear modulus in the continuum limit. In the units of current used in
the simulations this transition current is equal to $I_0=\sqrt{2}
b/(4\pi R_d) c_{66}/\tilde{c}_{66}$.

The vortex lattice is often considered as a conventional continuum
elastic medium characterized by its compressional $c_{11}$, shear
$c_{66}$, and tilt $c_{44}$ moduli, disregarding its discrete nature.
In the limit of very long wavelength distortions, however, the elastic
properties are governed by local elastic moduli, which strongly depend
on the strength of the magnetic field. Analytical results for the
local compressional, shear, and tilt moduli of a discrete vortex
lattice as a function of the lattice spacing $a$ are provided
elsewhere~\cite{MIG-00}.

We compute systematically the variation of $I_0$ with $N$ and $D$ (see
the inset of Fig.~\ref{fig5}).  When plotted as a function of the
lattice spacing $a\equiv\sqrt{2\pi D^2/\sqrt{3}N}$, the scattered data
can be collapsed into a single curve (Fig.~\ref{fig5}, main plot)
which follows the theoretical curve~\cite{MIG-00}
$c_{66}/\tilde{c}_{66}$.  When multiplied by an overall constant
$C=0.039\pm 0.003$ the curve provides a good fit of the data. This
constant is also in reasonable good agreement with its theoretical
estimate $C=\sqrt{2} b/(4\pi R_d)$ (where $b=a$ and the dipole extent
$R_d\sim 2-3 a$).

In conclusion, we have shown that the onset of plastic flow in a
superconducting disk occurs in close correspondence with the
nucleation and motion of new dislocations in the vortex lattice. Once
formed, dislocations glide parallel to their Burgers vector, releasing
the shear stress concentration. On increasing the nucleation rate,
dislocations arrange themselves into densely packed radial grain
boundaries that cross the entire disk and tend to glide in a
cooperative manner. The plastic response is characterized by a
non-linear I-V curve and by strong fluctuations.  These results have
been obtained for a two dimensional geometry, and they could change
when the thickness of the film is large. Our approach should be
relevant, however, for thin films of other self-assembled
nanostructures subject to the action of shearing forces
\cite{MUR-00,MIT-01,PER-01}, and possibly for granular media
\cite{VEJ-99}. Obviously, the precise value of the driving force
thresholds, signaling the onset of a nonlinear response and the
plastic deformation of the sample, would depend on the appropriate
physical parameters characterizing the interaction among the
constituent elements in each case.

\section*{Methods}
We consider a set of $N$ rigid vortices confined in a disk of radius
$D$. The equation of motion for each vortex $i$ at position ${\bf
r}_i$
\begin{equation}
\Gamma d{\bf r}_i/dt = \sum_j {\bf f}_{vv}({\bf r}_i - {\bf r}_j)+{\bf
f}_L({\bf r}_i),
\label{eq:vf}
\end{equation}
where $\Gamma$ is an effective viscosity. We choose as units of space
and time $\lambda$ and $t_0=\Gamma\lambda/A$ respectively, and we
measure the current $I$ in units of $\Phi_0/(2\pi c h \lambda A)$. The
$N$ vortices are confined inside the disk by the external magnetic
field and the sample edge barrier that we model by imposing an extra
normal force on the vortices of the form ${\bf f}_B=
-g\exp{[-(D-r)/r_0]}/r_0\hat{r}$, with $r_0= 0.1\lambda$ and
$g/A=1$. A similar force is also imposed at the inner wall close to
the disk center (at $r=r_0$), thus avoiding the singularity of the
Lorentz force at $r=0$.

The coupled Eqs.~(\ref{eq:vf}) for $i=1,...,N$ are integrated
numerically with an adaptive step size fifth-order Runge-Kutta method
with precision $10^{-6}$. We do not truncate the range of the
vortex-vortex interaction since this leads to spurious fluctuations
caused by the force discontinuities.  We study the response of the
system as a function of the applied current for different values of
$N$, ranging from $N=332$ to $N=2064$, and $D$ ($D=18 \lambda,\;
36\lambda, \; 72\lambda$).

\section*{Aknowledgements}
We thank G. Jung, M. Zaiser, R. Pastor-Satorras and J. S. Andrade
Jr. for useful remarks.  This work is supported by an Italy-Spain
Integrated Action.  MCM is supported by the Ministerio de Ciencia y
Tecnolog\'{\i}a (Spain). 

Correspondance and requests for materials should be addressed to
SZ. Supplementary Information accompanies the paper on
www.nature.com/naturematerials.

\end{document}